\documentclass{article}
\def\diracop{\not\negthinspace\partial}
\begin{document}
%%%%%%%%%%%%%%%%%%%%%%%%%%%%%%%%%%%%%%%%%%%%%%%%%%%%%%%%
\title{A Note on the Noncommutative Wess-Zumino Model
\footnote{hep-th/0009070, UFIFT-HEP-00-13}}
\author{Teparksorn Pengpan\footnote{email: pteparks@ratree.psu.ac.th}~  
and Xiaozhen Xiong\footnote{email: xiaozhen@phys.ufl.edu}}
\date{}
\maketitle
\centerline{\it Institute for Fundamental Theory,}
\centerline{\it Department of Physics, University of Florida}
\centerline{\it Gainesville FL 32611, USA}
\vskip 1cm
\begin{abstract}
We show that the noncommutative Wess-Zumino (NCWZ) Lagrangian with permutation terms 
in the interaction parts is renormalizable at one-loop level by only a wave function 
renormalization. When the non-commutativity vanishes, the logarithmic divergence 
of the wave function renormalization of the NCWZ theory 
is the same as that of the commutative one. Next the algebras of noncommutative 
field theories (NCFT's) are studied. From Neother currents, the field representation 
for the generators of NCFT's is extracted. Then based on this representation, 
the commutation relations between the generators are calculated for NCFT's. 
The symmetry properties of NCFT's inferred from these commutation relations are
discussed and compared with those of the commutative ones. 
\end{abstract}
\vskip 1cm
%\vspace\star {\fill}
\noindent PACS number(s): 11.10.-z, 11.10.Gh, 11.30.-j 12.60.Jv
\vskip 1cm
%\newline
%\newpage
\section{Introduction}

A deformation, an inverse of contraction (in the sense of Segal-Wigner-In\"{o}n\"{u} 
contraction), is one of the methods of generalization of a physical theory~\cite{Bayen}. 
The nondeformed theory can be recovered from the deformed one when taking 
a limit of deformation parameter to some value, e.g., nonrelativistic, 
classical physics, the nondeformed theory, is recovered from relativistic physics 
when taking the velocity of light $c\rightarrow \infty$, and from quantum physics 
when taking the Planck constant $h \rightarrow 0$. This naive concept has been applied 
to field theories on noncommutative (NC) spaces considered as deformations of flat Euclidean 
or Minkowski spaces. A product of fields on NC spaces can be 
expressed as a deformed product or star-product~\cite{Moyal,Zac}
 of fields on commutative spaces~\cite{Filk,SW,MRS}. 
Nevertheless, a question arises if the commutative field theories can be recovered
from their NC counter ones when non-commutativity $\Theta^{\mu\nu}\rightarrow 0$. 
At this moment, there is no conclusive answer to the question, and noncommutative 
field theories (NCFT's) 
must be investigated one by one.

An immediate question is the renormalizability of NCFT's. 
It was shown by Filk~\cite{Filk} that the NC complex scalar field theory 
has the same kind of divergences as the commutative one, 
and it was recently conjectured by Minwalla, 
Van Raamsdonk and Seiberg~\cite{MRS} that if a commutative theory is renormalizable, 
then the corresponding NC theory is also renormalizable, and if a commutative 
theory is not renormalizable, the corresponding NC theory is also not 
renormalizable. 
In \cite{CR}, the renormalizabity of the NC scalar field theory was studied, 
and the noncommutative Wess-Zumino (NCWZ) action on superspace was conjectured to be 
renormalizable. Later, in \cite{FL}, the deformation aspects of supersymmetric 
field theories were investigated, and the deformed Wess-Zumino model 
was again expected to be 
renormalizable. The superfield formulation of the NCWZ model is discussed and applied to 
the renormalization of the theory 
in \cite{CF} and \cite{AJH}. The detailed calculation of the 1PI diagram at one-loop 
has not been done yet. 

Another interesting issue is that NCFT's have nonlocal
interaction terms which explicitly break Lorentz invariance. However, the symmetry
must be broken in a particular way by the deformed star products of the fields
. Therefore, it is interesting to see how the Lorentz group is deformed in NCFT's.

In this paper, we investigate deformability and renormalizabity of the NCWZ theory 
on Minkowski space. In section 2, we review the wave function 
renormalization of the NC scalar $\Phi^4$ theory. In section 3, 
by adding permutation terms in the interaction part to preserve the 
supersymmetry transformations, we modify the original Wess-Zumino Lagrangian to be the NCWZ 
Lagrangian and investigate its renormalizability at one-loop order.
In section 4, we extract a representation for the algebras of NCFT's from  
Noether currents and calculate the commutation relations of the algebras.  
Then in the last section we compare our results with other works on  
NCFT's and comment on further research opportunities.

\section{The NC $\Phi^4$ theory}

To serve as an introduction to the renormalization of the NCWZ theory, 
let us review a result of the NC $\Phi^4$ theory in the four-dimensional space-time, 
which is described by
\begin{equation}
\mathcal{L} = \frac {1}{2} \partial_\mu \Phi\star\partial^\mu\Phi -\frac 12 m^2 \Phi\star\Phi 
- \frac \lambda{4!}\Phi\star\Phi\star\Phi\star\Phi. 
\label{phi4lagrangian}
\end{equation} 
As discussed in \cite{Filk} \cite{SW} and \cite{MRS}, 
under the integration, the star-product of fields does not 
affect the quadratic parts of the Lagrangian, whereas it makes the interaction part 
become nonlocal. Hence, Feynman rules in the momentum space of the NCFT  
are similar to those of the commutative one, except that the vertices of the NCFT 
are modified by a phase factor. For the Lagrangian (\ref{phi4lagrangian}),
the Feynman rule for the deformed vertex is
\begin{eqnarray}
& &-\frac i{3} \lambda
\left(\cos\frac 12(p_1\times p_2 + p_1\times p_3 + p_2\times p_3)\right.\nonumber\\
& &~~~~+ \cos\frac 12(p_1\times p_2 + p_1\times p_3 - p_2\times p_3)\nonumber\\
& &~~~~\left.+\cos\frac 12(p_1\times p_2 - p_1\times p_3 - p_2\times p_3)\right),
\end{eqnarray}
where $p_i$'s, $i=1\dots 4$, are the momenta coming out of the vertex, and 
$p_i\times p_j\equiv p_{i\mu}\Theta^{\mu\nu}p_{j\nu}$, where the non-commutativity 
$\Theta^{\mu\nu}$ is an anti-symmetric second rank tensor 
defined by $[q^\mu,q^\nu] = i\Theta^{\mu\nu}$. When $\Theta^{\mu\nu}\rightarrow 0$,
the deformed vertex becomes the non-deformed one. By using the above vertex, 
one yields a wave function renormalization of the scalar field $\Phi$ at one-loop order 
which has only one diagram as follows:
\begin{eqnarray}
\Gamma^{(\Phi\Phi)}(p^2) & = & i\frac \lambda6\int 
\frac{d^4k}{i(2\pi)^4}\frac{(2+\cos(p\times k))}{(k^2+m^2)}\nonumber\\
& = & i\frac{\lambda}{48\pi^2}\int_0^\infty \frac{d\alpha}{\alpha^2}e^{-i\alpha m^2}
\left(1+\frac 12 e^{-i\frac{\tilde{p}^2}{4\alpha}}\right)e^{\frac i{\Lambda^2\alpha}}\nonumber\\
& = & i\frac{\lambda}{48\pi^2}\left(\Lambda^2 - m^2ln(\frac{\Lambda^2}{m^2})\right)
+ i\frac{\lambda}{96\pi^2}\left(\Lambda_{eff}^2 - m^2ln(\frac{\Lambda_{eff}^2}{m^2})\right) 
+ \cdots.
\end{eqnarray}
The Schwinger parametrization technique to deal with the above integrations can be found in 
Itzykson and Zuber~\cite{IZu} and Hayakawa~\cite{Hayakawa}. In the second line, the term 
proportional to $\exp\left(-i\tilde{p}^2/4\rho\right)$, where
$\tilde{p}^\mu=\Theta^{\mu\nu}p_\nu$, is due to the nonplanar contribution, and the 
factor $\exp\left(i/\rho\Lambda^2\right)$ is introduced to regulate the small 
$\rho$ divergence in the planar contribution. Note that the nonplanar contribution 
is one-half of the planar one. In the third line, we keep only the divergent terms and 
the effective cutoff, $\Lambda_{eff}^2 = 1/\left(1/\Lambda^2+(-\tilde{p}^2)/4\right)$,
showing the mixing of ultraviolet (UV) divergence and Infrared (IR) singularity~\cite{MRS}. 
The above integration can also be done by using the dimensional regularization method, 
as shown in \cite{ABK1}. Renormalization of the theory at two loops is also discussed in 
detail in \cite{AM}. 

In the case that $\Phi$ is a complex scalar field, there are two ways in ordering the 
fields $\Phi$ and $\Phi^\ast$ in the quartic interaction $(\Phi^\ast\Phi)^2$. So, the 
general potential of the NC complex scalar field action is 
$$g_1\Phi^\ast\star\Phi\star\Phi^\ast\star\Phi+g_2\Phi^\ast\star\Phi^\ast\star\Phi\star\Phi.
$$
The potential is invariant under the global transformation, 
since the star product has nothing to 
do with the constant phase transformation. It was shown by Aref'eva, Belov and Koshelev
~\cite{ABK2} that the theory is not 
renormalizable for arbitrary values of $g_1$ and $g_2$, and is renormalizable at one-loop 
level only when $g_2=0$, or $g_1=g_2$.

\section{The NCWZ theory}

In this section, we focus on renormalization at one loop in the NCWZ theory. 
We modify the original WZ Lagrangian~\cite{WZ,IZ} 
to be the NCWZ Lagrangian by adding the permutation terms 
in the interaction part to preserve supersymmetry transformations.
Here, we follow the conventions by Sohnius~\cite{Soh}. The NCWZ model 
is described by the sum of the free off-shell Lagrangian and of the two invariants, 
\begin{equation}
\mathcal{L}_{tot} = \mathcal{L}_0 + \mathcal{L}_m + \mathcal{L}_g,
\label{NCWZ1}
\end{equation}
where
\begin{eqnarray}
\mathcal{L}_0 & = & \frac 12\left(\partial_\mu A \partial^\mu A 
+ \partial_\mu B \partial^\mu B +i\bar{\Psi} \diracop\Psi + F^2 + G^2\right),\\
\mathcal{L}_m & = & -m(FA+GB+\frac 12 \bar{\Psi} \Psi),\\
\mathcal{L}_g & = & -\frac g3\left(A\star A\star F-B\star B\star F+A\star B\star G+
\bar{\Psi}\star(A-\gamma_5 B)\star\Psi + \hbox{permutation terms} \right).
\label{NCWZ2}
\end{eqnarray}
The off-shell Lagrangians $\mathcal{L}_0$, $\mathcal{L}_m$ and $\mathcal{L}_g$ are separately invariant 
under the supersymmetry transformations:
\begin{equation}
\delta A=\bar{\alpha}\Psi,~~ 
\delta B=\bar{\alpha}\gamma_5\Psi,~~
\delta F=i\bar{\alpha}\diracop\Psi,~~
\delta G=i\bar{\alpha}\gamma_5\diracop\Psi,~~ 
\delta \psi=-(F+\gamma_5G)\alpha-i\diracop(A+\gamma_5B)\alpha, 
\end{equation}
where $\alpha$ and $\bar{\alpha}$ are the global infinitesimal Majorana spinor 
parameters. 

The Feynman rules in the momentum space can be extracted 
out directly from the Lagrangians (\ref{NCWZ1}). One gets as follows:
\begin{enumerate}
\item Propagators\newline 
The propagators of the fields and the mixed fields of the NC theory are 
the same as those of the commutative one.  
\item Deformed vertices
\begin{itemize}
\item  $-\frac g3(A\star A\star F + \hbox{permutation terms})$
$$-2ig\cos(\frac 12 p_{A_i}\times p_{A_o}).$$
\item  $\frac g3(B\star B\star F + \hbox{permutation terms})$
$$2ig\cos(\frac 12 p_{B_i}\times p_{B_o}).$$
\item $-\frac g3(A\star B\star G + \hbox{permutation terms})$
$$-2ig\cos(\frac 12 p_{A}\times p_{B}).$$
\item $-\frac g3(\bar{\Psi}\star A\star \Psi + \hbox{permutation terms})$
$$-2ig\cos(\frac 12 p_i\times p_o).$$
\item $\frac g3(\bar{\Psi}\star\gamma_5 B\star \Psi + \hbox{permutation terms})$
$$2ig\gamma_5\cos(\frac 12 p_i\times p_o).$$
\end{itemize}
\end{enumerate}
where the subscrips $i$ and $o$ label incoming and outgoing momenta.
The deformed vertices we obtain differ from the nondeformed ones by the factor, 
$\cos(\frac 12 p_i\times p_o)$. 

By using the above Feynman rules, one can calculate the one-loop UV divergent contributions 
to the 1PI 2-point and 3-point functions. The results are summarized as follows: 
\begin{enumerate}
\item Wave function renormalization
\begin{itemize}
\item Majorana field $\Psi$\newline
For the Majorana field, at one loop there are two diagrams. 
Their sum gives a contribution
\begin{eqnarray}
\Gamma^{(\bar{\Psi}\Psi)}(\not\negthinspace p) &=& -8ig^2\int \frac{d^4k}{i(2\pi)^4}
\cos^2(\frac 12 p\times k)\frac{\not\negthinspace k}{(k^2-m^2)((k+p)^2-m^2)}\nonumber\\
&=& i\not\negthinspace p\frac{g^2}{4\pi^2}\int_0^1d\alpha(1-\alpha)
\int_0^{\infty}\frac{d\rho}{\rho}e^{-\rho i\left(m^2-\alpha(1-\alpha)p^2\right)}
\left(1+e^{-\frac 1\rho\frac{i\tilde{p}^2}{4}}\right)e^{\frac 1\rho\frac {i}{\Lambda^2}}\nonumber\\
&=& i\not\negthinspace p\frac{g^2}{8\pi^2}\left(ln(\frac {\Lambda^2}{m^2})+
ln(\frac {\Lambda_{eff}^2}{m^2})\right)+\cdots.
\end{eqnarray}
\item Scalar fields $A,~B$ \newline
For each field, at one loop there are five diagrams. 
Their sum gives a contribution
\begin{eqnarray}
\Gamma^{(AA)}(p^2) = \Gamma^{(BB)}(p^2) &=& -8ig^2\int \frac{d^4k}{i(2\pi)^4}
\cos^2(\frac 12 p\times k)\frac{k\cdot p}{(k^2-m^2)((k+p)^2-m^2)}\nonumber\\
&=& ip^2\frac{g^2}{8\pi^2}\left(ln(\frac {\Lambda^2}{m^2})+
ln(\frac {\Lambda_{eff}^2}{m^2})\right)+\cdots.
\end{eqnarray}
\item Auxiliary fields $F,~G$ \newline
For the $F$ field, at one loop there are two diagrams. While, for the $G$ field, at 
one loop there is only one diagram. However, they give the same contribution
\begin{eqnarray}
\Gamma^{(FF)}(p^2) = \Gamma^{(GG)}(p^2) &=& 4ig^2\int \frac{d^4k}{i(2\pi)^4}
\cos^2(\frac 12 p\times k)\frac{1}{(k^2-m^2)((k+p)^2-m^2)} \nonumber\\
&=& i\frac{g^2}{8\pi^2}\left(ln(\frac {\Lambda^2}{m^2})+
ln(\frac {\Lambda_{eff}^2}{m^2})\right)+\dots.
\end{eqnarray}
\item Mixed fields
\begin{equation}
\Gamma^{(FA)}(p^2) = \Gamma^{(GB)}(p^2) = 0.
\end{equation}
\end{itemize}

Again, all the integrations can be done directly by using the Schwinger 
parametrization technique~\cite{IZu,Hayakawa}. The divergent terms of the one-loop 
corrections are the same for all the fields, whereas the finite terms of $\Gamma^{(FF)}$ 
and $\Gamma^{(GG)}$ are different from those of the others. However, all the finite terms 
are the functions of $p^2$ and $\tilde{p}^2$, and give finite contributions when $p =0$, 
i.e., there is no IR singularity. Note that in the NCWZ model the planar and 
nonplanar contributions have the same multiplicative factor, 
and when $\Theta^{\mu\nu}\rightarrow 0$, the right factor 
of the commutative Wess-Zumino model is retrieved.

\item Mass renormalizations
\begin{itemize}
\item Since, at one-loop $\Gamma^{(\bar{\Psi}\Psi)}(\not\negthinspace p)$ is proportional to 
only $\not\negthinspace p$, and both $\Gamma^{FA}$ and $\Gamma^{GB}$ are zero, the 
only mass renormaliztion is that due to the wave function renormalization. 
\end{itemize}
\item Vertex corrections
\begin{itemize}
\item $FA^2$, $FB^2$, $ABG$\newline
For each vertex, at one loop there are two diagrams, and they are added up to zero. So,
there is no correction for each vertex.
\item $\bar{\Psi} \Psi A$, 
$\bar{\Psi} \gamma_5 \Psi B$ \newline  
Similarly, there is no correction for each of these two vertices, since at one loop there 
are two diagrams, and they are added up to finite values.
\end{itemize}
\end{enumerate}

Just as in the $\Phi^4$ theory, the UV/IR mixing also appears in the NCWZ theory, 
which is the general consequence of 
the uncertainty relations between noncommutative coordinates~\cite{SW}. 
Renormalization in the NCWZ theory is very similar
to the commutative one. Compared with the ordinary
Wess-Zumino theory, the counter term for the wave function renormalization
reduces one-half, but the cancellations, in particular the absence
of mass and vertex corrections, persist due to supergauge invariance. 

\section{The Algebras of NCFT's}

In this section the algebras of NCFT's are studied. 
We'll follow the Noether's procedure to derive the conserved currents, from
which the generators are obtained, then the commutation relations between 
those generators are calculated. 

\subsection{Notations and Identities}

To facilitate the calculations involving NC fields star product, we introduce 
the following notations and list the useful identities.

Define an operator $\Delta$, which acts nontrivially on a scalar pair-product $(f,g)$ as, 
\begin{eqnarray}
\Delta(f,g) &\equiv& {\partial}_\mu f\tilde{\partial}^\mu g, \nonumber\\
\Delta^2(f,g) &=& \partial_\mu\partial_\nu f
\tilde{\partial}^\mu\tilde{\partial}^\nu g, \nonumber\\
\vdots & = & \vdots \nonumber\\
\Delta^n(f,g) &=& \underbrace{\partial_\mu\partial_\nu\cdots \partial_\rho}_n f
\underbrace{\tilde{\partial}^\mu\tilde{\partial}^\nu\cdots \tilde{\partial}^\rho}_n g,
\end{eqnarray}
where $\tilde{\partial}^\mu \equiv \frac i2\Theta^{\mu\nu}\partial_\nu$
\footnote{We include the factor $\frac i2$ here, slightly different from the definition 
in Section 2.}. 

With our definition, a star product between two scalar fields $A$ and $B$ can be written as
\begin{eqnarray}
A\star B & = & e^\Delta(A,B) \nonumber\\
& = & \left(1+\Delta+\frac{\Delta^2}{2!}+\frac{\Delta^3}{3!}+\cdots\right)(A,B)\nonumber\\
& = & AB + \partial_\mu\left(E(\Delta)(A,\tilde{\partial}^\mu B)\right),
\end{eqnarray}
where the operator $E(\Delta)$ is
\begin{equation}
E(\Delta) = \frac{e^\Delta-1}{\Delta} = \sum_{n=0}^{\infty}\frac{\Delta^n}{(n+1)!}.
\end{equation}

By using the above notations, we obtain some useful identities:
\begin{enumerate}
\item $B\star A  =  AB - \partial_\mu\left(E(-\Delta)(A,\tilde{\partial}^\mu B)\right).$
\item $[A,B]_{\star}  \equiv A\star B - B\star A = 2\partial_\mu\left(\frac{\sinh(\Delta)}{\Delta}(A,\tilde{\partial}^\mu B)\right).$
\item $\{A,B\}_{\star}  \equiv  A\star B + B\star A = 2AB + 2\partial_\mu\left(\frac{\cosh(\Delta)-1}{\Delta}(A,\tilde{\partial}^\mu B)\right).$
\item $(x_\rho A)\star B = x_\rho(A\star B) + A\star \tilde{\partial}_\rho B.$
\item $B\star (x_\rho A) = x_\rho(B\star A) - \tilde{\partial}_\rho B\star A.$
\item $[(x_\rho A),B]_\star =x_\rho[A,B]_\star + \{A,\tilde{\partial}_\rho B\}_\star.$
\item $[B,(x_\rho A)]_\star =x_\rho[B,A]_\star - \{A,\tilde{\partial}_\rho B\}_\star.$
\item $\{(x_\rho A),B\}_\star =x_\rho\{A,B\}_\star + [A,\tilde{\partial}_\rho B]_\star.$
\end{enumerate}

We assume $\theta^{0i} =0$ from now on for casuality and unitarity reasons~\cite{gomis}. The immediate
consequence is that non-commutativity will not introduce higher order time derivatives of
the fields in Lagrangian. 

\subsection{$\Phi^4$ theory}
Now let us calculate the Noether currents of 
the NC $\Phi^4$ theory following standard technique~\cite{Ramond} . Varying 
the Lagrangian (\ref{phi4lagrangian}), 
and using the above identites and also the equation of motion, one gets
\begin{equation}
\delta \int d^4x \mathcal{L}=\int d^4x
\partial_\mu\left(\frac 12 \{\partial^\mu\Phi,\delta_0\Phi\}_\star+\delta x^\mu\mathcal{L}
+\frac{\lambda}{12}\frac{\sinh(\Delta)}{\Delta}
(\Phi\star \Phi~,~\tilde{\partial}^\mu[\Phi,\delta_0\Phi]_\star)\right).
\end{equation}

Under an infinitesimal translation, $\delta x^\mu = g^{\mu\nu}\epsilon_\nu,
~\delta_0 \Phi = -\epsilon_\nu \partial^\nu \Phi$, one yields the energy-momentum tensor,
\begin{equation}
T^{\mu\nu} = \frac 12 \{\partial^\mu\Phi,\partial^\nu\Phi\}_\star - g^{\mu\nu}\mathcal{L}
+\frac{\lambda}{12}\frac{\sinh(\Delta)}{\Delta}
(\Phi\star \Phi~,~\tilde{\partial}^\mu[\Phi,\partial^\nu\Phi]_\star).
\label{energy-momentum-tensor}
\end{equation}
As explicitly seen, the energy-momentum tensor $T^{\mu\nu}$ is conserved 
since its divergence is zero. 
  
Under the infinitesimal Lorentz transformation, 
$\delta x^\mu = \epsilon^{\mu\nu}x_\nu = 
-\frac 12\epsilon^{\rho\sigma}(x_\rho g_\sigma^\mu-x_\sigma g_\rho^\mu),~
\delta_0 \Phi= 
\frac 12\epsilon^{\rho\sigma}(x_\rho\partial_\sigma\Phi-x_\sigma \partial_\rho \Phi)$, 
where $\epsilon^{\rho\sigma}$ is an anti-symmetric second rank tensor, 
one obtains a three-index current
\begin{eqnarray}
j_{\rho\sigma}^\mu & = & T_\rho^\mu x_\sigma 
+\frac 12[\partial_\rho\Phi,\tilde{\partial}_\sigma\partial^\mu\Phi]_\star 
+\frac{\lambda}{12}(\sinh(\Delta)/\Delta)^\prime
(\tilde{\partial}_\sigma(\Phi\star\Phi)~,
~\tilde{\partial}^\mu[\Phi,\partial_\rho\Phi]_\star) \nonumber\\
&  & -\frac{\lambda}{12}\frac{\sinh(\Delta)}{\Delta}
(\Phi\star\Phi~,~\tilde{g}_\sigma^\mu[\Phi,\partial_\rho\Phi]_\star+
\tilde{\partial}^\mu\{\tilde{\partial}_\sigma\Phi,\partial_\rho\Phi\}_\star)
-(\rho\leftrightarrow\sigma), 
\label{three-index current}
\end{eqnarray}
where $(\sinh(\Delta)/\Delta)^\prime=(\Delta \cosh(\Delta)-\sinh(\Delta))/\Delta^2$.
The divergence of the three-index current is not equal to zero due to the presence of the 
terms proportional to the non-commutativity $\Theta^{\mu\nu}$. However, note that the Noether currents of the 
commutative scalar field theory can be obtained by setting $\Theta^{\mu\nu}$ equal to zero. 

In the case of the commutative $\Phi^4$ theory, one yields 
the momentum and Hamiltonian generators 
from the energy-momentum tensor, and the angular momentum and boost generators 
from the three-index current 
\cite{Ramond}. These generators form the Poincar\'{e} algebra. For the NC 
$\Phi^4$ theory, one obtains its generators analogous to those of the commutative one,
\begin{eqnarray}
P^i & = & \int d^3x (\partial^i\Phi)\dot{\Phi} \equiv \int d^3x \mathcal{P}^i, \\
P^0 & = & \int d^3x \left(\frac 12(\dot{\Phi}^2+(\vec{\partial}\Phi)^2+m^2\Phi^2) 
+\frac{\lambda}{4!}\Phi^{\star 4}\right) \equiv \int d^3x \mathcal{P}^0,\\
M^{0i} & = & \int d^3x(x^0\mathcal{P}^i-x^i\mathcal{P}^0),\\
M^{ij} & = & \int d^3x(x^i\mathcal{P}^j-x^j\mathcal{P}^i).
\end{eqnarray}
The surface terms of $M^{0i}$ and $M^{ij}$ are dropped out. These generators generate 
the translational, rotational and boost transformations on $\Phi$. 

By using the quantization condition, 
$[\Phi(\vec x),\dot{\Phi}(\vec y)]=i\delta^{3}(\vec x-\vec y)$, 
one can easily obtain the following equal-time commutation relations:
\begin{eqnarray}
~[P^\mu,P^\nu] & = & 0, \label{trans}\\
~[M^{ij},M^{kl}] & = & i(\eta^{il}M^{jk}+\eta^{jk}M^{il}-\eta^{ik}M^{jl}-\eta^{jl}M^{ik}), \\
~[M^{ij},P^k] & = & i(\eta^{jk}P^i-\eta^{ik}P^j), \\
~[M^{0i},P^j] & = & i\eta^{ij}P^0.
\end{eqnarray}

The above commutation relations of the NC $\Phi^4$ theory are the same as 
those of the commutative one. In particular, ~(\ref{trans}) verifies that the 
NC $\Phi^4$ Lagrangian has translational invariance and 
the translation generator 
$P^\mu$ is conserved. But the following commutation relations have some additional 
terms proportional to $\Theta^{\mu\nu}$, 
due to the symmetry-breaking term $\frac {\lambda }{4!} \Phi^{*4}$,
\begin{eqnarray}
~[M^{0i},P^0] & = & -i\eta^{00}P^i-i\frac{\lambda}{4!}
\int d^3x\{\dot{\Phi},[\Phi^{\star 2},\tilde{\partial}^i\Phi]_\star\}_\star , 
\label{boost}\\
~[M^{ij},P^0] & = & -i\frac{\lambda}{3!}\int d^3x x^i(\partial^j\Phi)\Phi^{\star 3} + 
(i\leftrightarrow j), 
\label{rot}\\
~[M^{0i},M^{0j}] & = & -i\eta^{00}M^{ij}+i\frac{\lambda}{4!}
\int d^3x \left(x^j\{\dot{\Phi},[\Phi^{\star 2},\tilde{\partial}^i\Phi]_\star\}_\star 
-(i\leftrightarrow j)\right), \\
~[M^{0i},M^{jk}] & = & i(\eta^{ij}M^{0k}-\eta^{ik}M^{0j})
-i\frac{\lambda}{4!} \int d^3x x^i \nonumber\\
&&\times\left([\partial^k\Phi ,\tilde{\partial}^j\Phi^{\star 3}]_\star \right. 
+\Phi^{\star 2}\star\partial^k\Phi\star\tilde{\partial}^j\Phi 
-\tilde{\partial}^j\Phi^{\star 2}\star\partial^k\Phi\star\Phi \nonumber \\
&& \left.+\Phi\star\partial^k\Phi\star\tilde{\partial}^j\Phi^{\star 2} 
-\tilde{\partial}^j\Phi\star\partial^k\Phi\star\phi^{\star 2}-(j \leftrightarrow k)\right).
\end{eqnarray}

The eqns~(\ref{boost}) and (\ref{rot}) explicitly show that the Lorentz generators 
are not conserved
in the theory, and all the deformation terms are directly proportional to $\Theta^{\mu\nu}$.

\subsection{Wess-Zumino model}
For the NCWZ model, one start from an on-shell Lagrangian
analogous to the commutative one~\cite{Soh},
\begin{eqnarray}
\mathcal{L} & = & \frac 12(\partial_\mu A\partial^\mu A-m^2A^2)
+\frac 12(\partial_\mu B\partial^\mu B-m^2B^2)
+\frac 12(i\bar{\Psi}\diracop\Psi-m\bar{\Psi}\Psi) \nonumber \\
 & & -mgA(A^{\star 2}+B^{\star 2})-mgB(A\star B+B\star A) \nonumber \\
 &&-g(A\bar{\Psi}\star\Psi-B\bar{\Psi}\star\gamma_5\Psi)
 -\frac 12g^2(A-iB)^{\star 2}(A+iB)^{\star 2} \\
&=&\frac 12(\partial_{\mu} \phi\partial^{\mu} \bar{\phi}-m^2\phi\bar{\phi})
+\frac 12(i\psi\sigma^{\mu}\partial_{\mu}\bar{\psi}
+i\bar{\psi}\bar{\sigma}^\mu\partial_\mu \psi-m\bar{\psi}\bar{\psi}-m\psi\psi) \nonumber \\
 & & -\frac 12 mg(\phi\bar{\phi}^{\star 2}+\bar{\phi}\phi^{\star 2})
 -g(\phi\bar{\psi}\star\bar{\psi}+\bar{\phi}\psi\star\psi)
 -\frac 12 g^2\phi^{\star 2}\bar{\phi}^{\star 2}.
\end{eqnarray}
where $\phi\equiv A-iB,\bar{\phi}\equiv A+iB $, and $\psi,\bar{\psi}$ are the Weyl components 
of the Majorana field $\Psi$,
 following the notations and conventions by Bailin and Love~\cite{BL}. 

Following the similar procedure as done in the $\phi^4$ theory, the variation of 
the Lagrangian under the infinitesimal Poincar\'{e} and supergauge transformations 
yields the generators as,   
\begin{eqnarray}
P^i & = & \int d^3x\left(\frac 12\partial^i \phi\dot{\bar{\phi}}+\frac 12\dot{\phi}\partial^i\bar{\phi}
+i\bar{\psi}\bar{\sigma}^0\partial^i\psi\right)\equiv \int d^3x \mathcal{P}^i, \\
P^0 & = & \int d^3x\left(\frac 12(\dot{\phi}\dot{\bar{\phi}}
+\partial^i\phi\partial^i\bar{\phi}
+m^2\phi\bar{\phi})
+\frac 12(i\bar{\psi}\bar{\sigma}^i\partial^i\psi
+i\psi\sigma^i\partial^i\bar{\psi}
+m\psi\psi+m\bar{\psi}\bar{\psi}) \right.\nonumber\\
&&\left.+\frac 12 mg(\phi\bar{\phi}^{*2}+\bar{\phi}\phi^{*2})
  +g(\phi\bar{\psi}\star\bar{\psi}+\bar{\phi}\psi\star\psi)
  +\frac 12 g^2\phi^{*2}\bar{\phi}^{*2}\right)
  \equiv \int d^3x\mathcal{P}^0, \\ 
M^{0i} & = & \int d^3x\left(x^0\mathcal{P}^i-x^i\mathcal{P}^0\right),\\
M^{ij} & = & \int d^3x\left(x^i\mathcal{P}^j-x^j\mathcal{P}^i\right),\\
\chi Q & = & \chi\int d^3x\left(\dot{\phi}\psi-2\partial_i\phi\sigma^{0i}\psi
+im\phi\sigma^{0}\bar{\psi}+ig\phi^{\star 2}\sigma^{0}\bar{\psi}\right),\\
\bar{\chi}\bar{Q} & = & \bar{\chi}\int d^3x\left(\dot{\bar{\phi}}\bar{\psi}
-2\partial_i\bar{\phi}\bar{\sigma}^{0i}\bar{\psi}
+im\bar{\phi}\bar{\sigma}^{0}\psi+ig\bar{\phi}^{\star 2}\bar{\sigma}^{0}\psi\right)
~=~(\chi Q)^\dag,
\end{eqnarray}
where $\chi$ is an arbitrary Majorana spinor parameter.

In the case of the commutative Wess-Zumino model, the analogs of the above generators are 
those of the Poincar\'{e} algebra and supercharge, which form the $N=1$ super-Poincar\'{e} 
algebra. With the representations obtained here in the NCWZ model, 
one can calculate the commutation relations between those generators, 
\begin{equation}
[P^\mu,P^\nu] =  0,\label{wztrans}
\end{equation}
\begin{equation}
[M^{ij},M^{kl}]  = i(\eta^{il}M^{jk}+\eta^{jk}M^{il}-\eta^{ik}M^{jl}-\eta^{jl}M^{ik}),
\end{equation}
\begin{equation}
[M^{ij},P^k] = i(\eta^{jk}P^i-\eta^{ik}P^j),
\end{equation}
\begin{equation}
[M^{0i},P^j]  =  i\eta^{ij}P^0,\label{special}
\end{equation}

The above commutation relations are exactly the same as those obtained in the NC 
$\Phi^4$ theory, which suggests the generality of such relations for all NCFT's. 
In particular, (\ref{wztrans}) verifies the translational invariance
of the theory. Equation (\ref{special}) is a little surprising. The calculation of
it in any way involves the NC interaction terms. Nevertheless it's 
true for both NCFT's.

Other commutation relations are,
\begin{equation}
[\chi Q,\zeta Q] = [\bar{\chi} \bar{Q},\bar{\zeta} \bar{Q}] = 0,\label{QQ}
\end{equation}
\begin{equation}
[\chi Q,\bar{\zeta} \bar{Q}] = 2\chi\sigma^\mu\bar{\zeta}P_\mu,\label{QQbar}
\end{equation}
\begin{equation}
[P^\mu, \chi Q] = 0,\label{PQ}
\end{equation}
\begin{equation}
[M^{ij}, \chi Q] = -i\chi \sigma^{ij}Q,
\end{equation}
\begin{equation}
[M^{ij}, \bar{\chi} \bar{Q}] = -i\bar{\chi} \bar{\sigma}^{ij}\bar{Q}.
\end{equation}
All the above relations are exactly the same as 
those of the commutative Wess-Zumino model. In particular, one 
finds the supercharge generators, $Q$ and $\bar{Q}$, and the translation
generators $P^\mu$'s form a close algebra, and the supercharge generators 
are conserved. 

The rest commutation relations have additional
terms proportional to $\Theta^{\mu\nu}$, including the similar ones as appears 
in the NC $\Phi^4$ theory,  
\begin{eqnarray}
[M^{0i},P^0] & = & -i\eta^{00}P^i-\int d^3x \left(\frac i2 
mg([\phi,\tilde{\partial}^i\bar{\phi}]_\star \dot{\bar{\phi}}
   +\dot{\phi}[\bar{\phi},\tilde{\partial}^i\phi]_\star) 
\right.
+mg([\bar{\phi},\tilde{\partial}^i\psi]_\star\sigma^0\bar{\psi}
      +[\phi,\tilde{\partial}^i\bar{\psi}]_\star\bar{\sigma}^0\psi)\nonumber\\
&& -2ig([\bar{\phi},\tilde{\partial}^i\psi]_\star\sigma^{0l}\partial_l\psi
      +[\phi,\tilde{\partial}^i\bar{\psi}]_\star\bar{\sigma}^{0l}\partial_l\bar{\psi})
   +\frac i2 g^2(\dot{\phi}[\bar{\phi}^{\star 2},\tilde{\partial}^i\phi]_\star
               +[\phi^{\star 2},\tilde{\partial}^i\bar{\phi}]_\star\dot{\bar{\phi}})
\nonumber\\
&&\left.  +g^2([\phi,\tilde{\partial}^i\psi]_\star\{\bar{\phi},\psi\}_\star
       -\{\phi,\bar{\psi}\}_\star[\bar{\phi},\tilde{\partial}^i\psi]_\star)
 \right),
\end{eqnarray}
\begin{eqnarray}
[M^{ij},P^0] & = & \int d^3x \left(
\frac i2 mg([\partial^i\phi,\tilde{\partial}^j\phi]_\star\bar{\phi}
            +[\partial^i\bar{\phi},\tilde{\partial}^j]_\star\phi)
+ig([\partial^i\bar{\psi},\tilde{\partial}^j\bar{\psi}]_\star\phi
   +[\partial^i\psi,\tilde{\partial}^j\psi]_\star\bar{\phi})\right.\nonumber\\
&&\left.+\frac i2 g^2([\partial^i\phi,\tilde{\partial}^j\phi]_\star\bar{\phi}^{\star 2}
             +[\partial^i\bar{\phi},\tilde{\partial}^j\bar{\phi}]_\star\phi^{\star 2})
\right)-(i\leftrightarrow j),
\end{eqnarray}
\begin{eqnarray}
[M^{0i},M^{0j}]&=&-i\eta^{00}M^{ij}+\int d^3x \left(
mgx^i(\psi\sigma^0[\phi,\tilde{\partial}^j\bar{\psi}]_\star
     +[\bar{\phi},\tilde{\partial}^j\psi]_\star\sigma^0\bar{\psi})\right.\nonumber\\
&&-\frac i2 mgx^i([\phi,\tilde{\partial}^j\bar{\phi}]_\star\dot{\bar{\phi}}
                  +\dot{\phi}[\bar{\phi},\tilde{\partial}^j\phi]_\star)
  +2igx^i([\phi,\tilde{\partial}^j\bar{\psi}]_\star\bar{\sigma}^{0l}\partial_l\bar{\psi}
        +[\bar{\phi},\tilde{\partial}^j\psi]_\star\sigma^{0l}\partial_l\psi)\nonumber\\                  
&&\left.
  -\frac i2 g^2x^i([\phi^{\star 2},\tilde{\partial}^j\bar{\phi}]_\star\dot{\bar{\phi}}
                +\dot{\phi}[\bar{\phi}^{\star 2},\tilde{\partial}^j\phi]_\star)
       +g^2\{x^i\phi,\bar{\psi}\}_\star\{\psi,x^j\bar{\phi}\}_\star
       -(i\leftrightarrow j)\right),         
\end{eqnarray}
\begin{eqnarray}
\lefteqn{[M^{0i},M^{jk}]~=~}\nonumber\\
&&i(\eta^{ij}M^{0k}-\eta^{ik}M^{0j})-\int d^3x \left(
\frac i2 mgx^i(\bar{\phi}[\partial^k\phi,\tilde{\partial}^j\phi]_\star
+\phi[\partial^k\bar{\phi},\tilde{\partial}^j\bar{\phi}]_\star)\right.\nonumber\\
&& \left.+igx^i(\phi[\partial^k\bar{\psi},\tilde{\partial}^j\bar{\psi}]_\star
         +\bar{\phi}[\partial^k\psi,\tilde{\partial}^j\psi]_\star)
   +\frac i2 g^2x^i([\partial^k\phi,\tilde{\partial}^j\phi]_\star\bar{\phi}^{\star 2}
                   +[\partial^k\bar{\phi},\tilde{\partial}^j\phi]_\star)
  -(j\leftrightarrow k)\right),
\end{eqnarray}
 
and also the transformations of the supercharge generators under the Lorentz boosts, 
\begin{eqnarray}
[M^{0i}, \chi Q] & = & -i\chi \sigma^{0i}Q
+\int d^3x\left(g[\dot{\phi},\phi]_\star\chi\sigma^0\tilde{\partial}^i\bar{\psi}
+g[\phi,\tilde{\partial}^i\partial_l(\phi)]\chi\sigma^l\bar{\psi}\right.\nonumber\\
& & \left.-2ig\psi\star\psi\chi\psi
+img[\phi,\bar{\phi}]_\star\chi\tilde{\partial}^i\psi
+ig^2[\phi^{\star 2},\bar{\phi}]_\star\chi\tilde{\partial}^i\psi\right),
\end{eqnarray}
\begin{eqnarray}
[M^{0i}, \bar{\chi} \bar{Q}] & = & -i\bar{\chi} \bar{\sigma}^{0i}\bar{Q}
+\int d^3x\left(g[\dot{\bar{\phi}},\bar{\phi}]_\star\bar{\chi}\bar{\sigma}^0\tilde{\partial}^i\psi
+g[\bar{\phi},\tilde{\partial}^i\partial_l\bar{\phi}]\bar{\chi}\bar{\sigma}^l\psi
\right.\nonumber\\
& & \left.-2ig\bar{\psi}\star\bar{\psi}\bar{\chi}\bar{\psi}
+img[\bar{\phi},\phi]_\star\bar{\chi}\tilde{\partial}^i\bar{\psi}
+ig^2[\bar{\phi}^{\star 2},\phi]_\star\bar{\chi}\tilde{\partial}^i\bar{\psi}\right).
\end{eqnarray}
To simplify the expression, we reorder the conjugate fields
on the right hand side of the above equations, which induces extra infinite contant terms 
not explicitly shown here. 

In summary,  The commutation relations of the Lorentz
rotation and boost generators generally have additional terms 
compared with those of the Poincar\'{e} or super-Poincar\'{e} algebras. 
Nevertheless, the results are not surprising, since NCFT's 
indeed violate the Lorentz invariance. Other commutation relations
verify certain symmetries preserved by NCFT's, such as 
the translational and supergauge invariance. In the limit of 
$\Theta^{\mu\nu}\rightarrow 0$, the Poincar\'{e} or Super-Poincar\'{e} algebra is retrieved.

\section{Conclusions} 
In this paper we first construct a NCWZ Lagrangian, 
from which the Feynman rules are extracted, then the one-loop UV divergent corrections  
to the 1PI 2-point functions are explicitly calculated and the renormalization of the 
theory at one-loop are studied. 
We found that Girotti and collaborators~\cite{GGRS} studied the NCWZ 
theory by using the Lagrangian similar to ours without using the 
entirely permutation terms in the interaction parts. However, we arrive at the same 
conclusion, i.e. the NCWZ model is renormalizable by only 
a wave function renormalization, as expected by Ferrara and Lled\'{o}~\cite{FL}. 
But, our calculations explicitly show that 
the UV/IR mixing still exists in the divergent terms and the renormalization of the wave 
function of the commutative theory can be recovered by setting $\Theta^{\mu\nu}$ equal to zero. 

Next we turn to the algebras of the NC $\phi^4$ and 
Wess-Zumino theory. From Noether currents we extract a representation of 
the translation, Lorentz and supercharge generators, 
which is what Dirac called `fundamental quantities'
~\cite{Dirac} for NCFT's. The commutation relations of those
quantities are calculated based on this representation. 

The NCFT has non-local interaction terms,
which explicitly break the Lorentz invariance, but still preserve the translational and
supergauge invariance. It's found that in the NCFT the translation 
and supercharge generators form the same algebra as in the commutative theory. 
But, the commutation relations of the Lorentz generators, or between 
the Lorentz generators and the translation or supercharge generators, generally 
have extra terms proportional to the non-commutativity $\Theta^{\mu\nu}$. 
In addition to that, there are also other 
interesting commutation relations, such as $[M^{0i},P^j]=i\eta^{ij}P^0$, 
still hold true in the NC case. 

The role of the representations for the algebras is not clear yet. 
Since those representations for the fundamental quantities
could also construct a theory of a dynamical system~\cite{Dirac}
, questions, like `Is the theory so constructed exactly equivalent to the 
theory with the original Lagrangian?', `Can the extra terms, which appear in the 
commutation relations of the non-invariant fundamental quantities, 
actually be expressed by other generators and thus all the generators form 
a deformed Lorentz algebra?', have yet to be answered. 

\vskip 1cm
\centerline{\bf Acknowledgments}
We would like to thank Professor P. Ramond and Dr. K. Bering for many stimulating and 
informative discussions. This work was supported by the Department of Energy 
under grant DE-FG02-97ER-41029.

%\newpage

\end{document}